\documentclass[preprint2]{aastex}

  \usepackage{subfigure} 
\usepackage{graphicx}
\usepackage{epsf}

\newcommand{\sm}[1]{\mbox{{\scriptsize #1}}}

\newcommand{\bef}{\begin{figure}}
\newcommand{\eef}{\end{figure}}

\def\eps@scaling{0.95}

\def\showone#1{
  \centering
  \leavevmode
  \epsfxsize=\eps@scaling\linewidth
  \epsfbox{#1.eps}
}

\def\showtwover#1#2{
  \centering
  \leavevmode
  \epsfxsize=\eps@scaling\linewidth
  \epsfbox{#1.eps} \hfil
  \epsfxsize=\eps@scaling\linewidth
  \epsfbox{#2.eps}
}

\def\showthreeover#1#2#3{
  \centering
  \leavevmode
  \epsfxsize=\eps@scaling\linewidth
  \epsfbox{#1.eps} \hfil
  \epsfxsize=\eps@scaling\linewidth
  \epsfbox{#2.eps} \hfil
  \epsfxsize=\eps@scaling\linewidth
  \epsfbox{#3.eps}
}

\def\showfourover#1#2#3#4{
  \centering
  \leavevmode
  \epsfxsize=\eps@scaling\linewidth
  \epsfbox{#1.eps} \hfil
  \epsfxsize=\eps@scaling\linewidth
  \epsfbox{#2.eps} \hfil
  \epsfxsize=\eps@scaling\linewidth
  \epsfbox{#3.eps} \hfil
  \epsfxsize=\eps@scaling\linewidth
  \epsfbox{#4.eps}
}

\def\epstwo@scaling{0.46}

\def\showtwo#1#2{
  \centering
  \leavevmode
  \epsfxsize=\epstwo@scaling\linewidth
  \epsfbox{#1.eps} 
  \epsfxsize=\epstwo@scaling\linewidth
  \epsfbox{#2.eps}
}

\def\epsthree@scaling{0.28}
\def\showthree#1#2#3{
  \centering
  \leavevmode
  \epsfysize=\epsthree@scaling\textwidth 
  \epsfbox{#1.eps} 
  \epsfysize=\epsthree@scaling\textwidth 
  \epsfbox{#2.eps}
  \epsfysize=\epsthree@scaling\textwidth 
  \epsfbox{#3.eps}
}

\def\epstwo@scaling{0.44}

\def\showfour#1#2#3#4{
  \centering
  \leavevmode
  \epsfxsize=\epstwo@scaling\linewidth
  \epsfbox{#1.eps} \hfil
  \epsfxsize=\epstwo@scaling\linewidth
  \epsfbox{#2.eps} \hfil
  \epsfxsize=\epstwo@scaling\linewidth
  \epsfbox{#3.eps} \hfil
  \epsfxsize=\epstwo@scaling\linewidth
  \epsfbox{#4.eps}
}

\def\showsix#1#2#3#4#5#6{
  \centering
  \leavevmode
  \epsfxsize=\epstwo@scaling\linewidth
  \epsfbox{#1.eps} \hfil
  \epsfxsize=\epstwo@scaling\linewidth
  \epsfbox{#2.eps} \hfil
  \epsfxsize=\epstwo@scaling\linewidth
  \epsfbox{#3.eps} \hfil
  \epsfxsize=\epstwo@scaling\linewidth
  \epsfbox{#4.eps} \hfil
  \epsfxsize=\epstwo@scaling\linewidth
  \epsfbox{#5.eps} \hfil
  \epsfxsize=\epstwo@scaling\linewidth
  \epsfbox{#6.eps}
}

\newcommand{\befone}{
  \begin{figure*}
  \centering
  \begin{minipage}{\textwidth}
  }
\newcommand{\eefone}{\end{minipage}\end{figure*}}

\newcommand{\HzI}{$\mathrm{H}_2$\ }

\newcommand{\HeHII}{$\mathrm{HeH}^+$\ }

\newcommand{\COI}{$\mathrm{CO}$\ }

\newcommand{\CII}{$\mathrm{C}^+$\ }

\newcommand{\OI}{$\mathrm{O}$\ }

\newcommand{\HzId}{$\mathrm{H}_2$}

\newcommand{\COId}{$\mathrm{CO}$}

\newcommand{\OId}{$\mathrm{O}$}

\shorttitle{The prospects of finding the first quasars}
\shortauthors{Schleicher et al.}

\begin{document}


\title{The prospects of finding the first quasars in the universe}


\author{Dominik R. G. Schleicher\altaffilmark{1}, Marco Spaans\altaffilmark{2}, Ralf~S.~Klessen\altaffilmark{1}}
\email{dschleic@ita.uni-heidelberg.de, spaans@astro.rug.nl, rklessen@ita.uni-heidelberg.de}

\altaffiltext{1}{Zentrum f\"ur Astronomie der Universit\"at Heidelberg, Institut f\"ur theoretische Astrophysik, Albert-Ueberle-Str. 2, D-69120 Heidelberg, Germany}
\altaffiltext{2}{Kapteyn Astronomical Institute, University of Groningen, P.O. Box 800, 9700 AV, Groningen, the Netherlands}

\begin{abstract}
We study the prospects of finding the first quasars in the universe with ALMA and JWST. For this purpose, we derive a model for the high-redshift black hole population based on observed relations between the black hole mass and the host galaxy. We re-address previous constraints from the X-ray background with particular focus on black hole luminosities below the Eddington limit as observed in many local AGN. For such luminosities, up to $20\%$ of high-redshift black holes can be active quasars. We then discuss the observables of high-redshift black holes for ALMA and JWST by adopting NGC~1068 as a reference system. We calculate the expected flux of different fine-structure lines for a similar system at higher redshift, and provide further predictions for high-$J$~\COI lines. We discuss the expected fluxes from stellar light, the AGN continuum and the Lyman $\alpha$ line for JWST. Line fluxes observed with ALMA can be used to derive detailed properties of high-redshift sources. We suggest two observational strategies to find potential AGN at high redshift and estimate the expected number of sources, which is between $1-10$ for ALMA with a field of view of $\sim(1')^2$ searching for line emission and $100-1000$ for JWST with a field of view of $(2.16')^2$ searching for continuum radiation. We find that both telescopes can probe high-redshift quasars down to redshift $10$ and beyond, and therefore truely detect the first quasars in the universe.
\end{abstract}

\section{Introduction}
After the first detection of supermassive black holes at $z\sim6$ \citep{Fan01}, there has been a strong effort to find more sources at high redshift and to study their properties. Further high-$z$ quasi-stellar objects (QSOs) have been found in surveys like the Sloan Digital Sky Survey (SDSS)  \citep{Fan01,Carballo06,Cool06,Fan06,Shen07,Inada08}, the Canada-France high-redshift quasar survey \citep{Willott07}, the UKIDSS Large Area Survey \citep{Venemans07} and the Palomar-QUEST survey \citep{Bauer06}. They have been studied by follow-up observations with Chandra \citep{Shemmer06}, the Subaru Telescope \citep{Goto07}, with IR spectroscopy \citep{Juarez07}, near-IR spectroscopy \citep{Jurez08}, mid-IR observations \citep{Stern07, Martinez08}, Swift observations \citep{Sambruna07}, VLBA observations \citep{Momjian07}, the Very Large Array as well as the Very Long Baseline Array \citep{Momjian08}. Some of their host galaxies have also been detected in CO and submm emission \citep{Omont96, Carilli02, Walter04, Weiss07, Riechers08b, Riechers08a}.

These observations furthermore inferred a number of relations between the mass of the central black hole and the properties of their host galaxy. The Magorrian relation connects the velocity dispersion and the mass of the galactic bulge with the black hole mass \citep{Magorrian98, Ferrarese00, Gebhardt00,Graham01,Merritt01,Tremaine02,Haering04,Peterson08,Somerville08}. Quasars with black hole masses of the order $10^9\,M_\odot$ are observed with supersolar metallicities even at $z\sim6$ \citep{Pentericci02}, which is indicated in further studies as well \citep{Freudling03,Maiolino03,Dietrich03a,Dietrich03b}. In the nearby universe, correlations have been found between black hole mass and metallicity \citep{Warner03, Kisaka08}, with a mild slope of $0.38\pm0.07$. This is plausible, as it is generally recognized that the overall mass of a galaxy is correlated with its metallicity \citep{Faber73,Zaritsky94,Jablonka96,Trager00}. In this paper, we will in general assume that the black hole mass is a good indicator for the evolutionary stage of the quasar host galaxy, in particular with respect to the Magorrian relation and the expected metallicity.

In the local universe, active galactic nuclei (AGN) can be studied in much more detail and with higher resolution. An example that has been studied particularly well, in different wavelengths and with high resolution is the AGN NGC~1068. Its direct X-ray emission is shielded through absorption along the line of sight, but there is evidence for reflected X-ray photons that are scattered into the line of sight \citep{Pounds06}. Its molecular disk has been studied through emission from different species, in particular molecular \COI and fine-structure lines \citep{Schinnerer00, Galliano03, Galliano05, Spinoglio05, Davies07, Davies07b, Poncelet07}. The continuum emission and the distribution of dust has been measured in further studies \citep{Krips06, Mason06, Rhee06, Tomono06, Howell07, Hoenig08}. Such observations in particular allow to probe stellar and gaseous dynamics and the galactic gravitational potential \citep{Gerssen06, Emsellem06, Das07}. Observations with the Green Bank Telescope (GBT) provide evidence for water maser activity \citep{Greenhill95, Greenhill96, Kondratko06}. The inner radio jet can be detected with observations at $7$~mm wavelength \citep{Cotton08}. Given this amount of detail, NGC~1068 thus provides an excellent test case for the inner structure of active galaxies. 

The formation of quasars and their supermassive black holes is still one of the unresolved riddles of structure formation and cosmology. The simplest scenarios assume that they have grown from the remnants of the first stars, which are believed to be very massive \citep{Abel02, Bromm04, Glover05}, and whose black hole remnants could grow further by accretion. Such a scenario however has problems, as the remnants of the first stars typically do not end up in the most massive quasars at redshift $z\sim6$ \citep{Trenti08}, and radiative feedback from the stellar progenitor can delay accretion as well \citep{Johnson07, Alvarez08, Milosavljevic08}. In case of Eddington accretion, seed black holes of $\sim10^5\,M_\odot$ are required in order to grow to the observed supermassive black holes at $z\sim6$ \citep{Shapiro05}. 

Recently, it has also been discussed whether the first stars in the early universe were powered by dark matter annihilation rather than nuclear fusion \citep{Spolyar08, Iocco08}. Such stars could reach masses of the order $1000\ M_\odot$ \citep{FreeseBodenheimer08, IoccoBressan08} and were considered as possible progenitors for the first supermassive black holes. The evolution of such stars on the main sequence has been calculated by \citet{Taoso08} and \citet{Yoon08}. However, it was shown that such stellar models are highly constrained by the observed reionization optical depth \citep{SchleicherBanerjee08a, SchleicherBanerjee08c}. Also, we note that such seeds would still require super-Eddington accretion to grow to the observed supermassive black holes at $z\sim6$.

Therefore, alternative scenarios have been considered that lead to the formation of massive seed black holes by direct collapse \citep{Eisenstein95, Koushiappas04, Begelman06, Spaans06, Dijkstra08}. However, for such scenarios, the situation is also not entirely clear. For instance, \citet{Lodato06} argue that the gas in such halos should fragment if \HzI cooling is efficient, and \citet{Omukai08} and \citet{Jappsen08} suggest that a non-zero metallicity will lead to fragmentation as well. Simulations by \citet{Clark08} support these conjectures. 

The black hole population at high redshift is further constrained by observations of the soft X-ray background. These constraints indicate that the population of high-redshift quasars was not sufficient to reionize the universe \citep{Dijkstra04}, and suggest an upper limit to the black hole density of $\sim4\times10^4\,M_\odot\,\mathrm{Mpc}^{-3}$ \citep{Salvaterra05}. The latter depends in particular on the adopted spectra and luminosity of the black hole, as well as their duty cycles, and it has been argued that larger black hole populations are conceivable as well \citep{Zaroubi07}. It is therefore important to probe the number density of black holes and AGN activity as a function of redshift by direct observations.

In this work, we explore the possibility of searching for the first quasars in the universe with ALMA \footnote{http://www.eso.org/sci/facilities/alma/index.html} and JWST \footnote{http://www.stsci.edu/jwst/overview/}. ALMA is particularly relevant in light of recent predictions on the observability of the first miniquasars \citep{Spaans08} and high-redshift Lyman $\alpha$ galaxies \citep{Finkelstein08}, while it is one of the main scientific goals of JWST to find the first galaxies at high redshift. In \S \ref{population}, we introduce a model for the black hole population which is based on observed properties and correlations between black holes and their host galaxy. In \S \ref{constraint}, we discuss how the number of active black holes is constrained by the unresolved X-ray background. In \S \ref{ngc1068}, we analyze the observed properties of NGC~1068 and derive further predictions for very high-$J$~\COI lines. In \S \ref{observables}, we use these information to derive the relevant observables for ALMA and JWST and predict the number of detectable sources as a function of redshift. Further discussion and outlook is provided in \S \ref{discussion}.

\section{The high-redshift black hole population}\label{population}
\begin{figure}[t]
\includegraphics[scale=0.45]{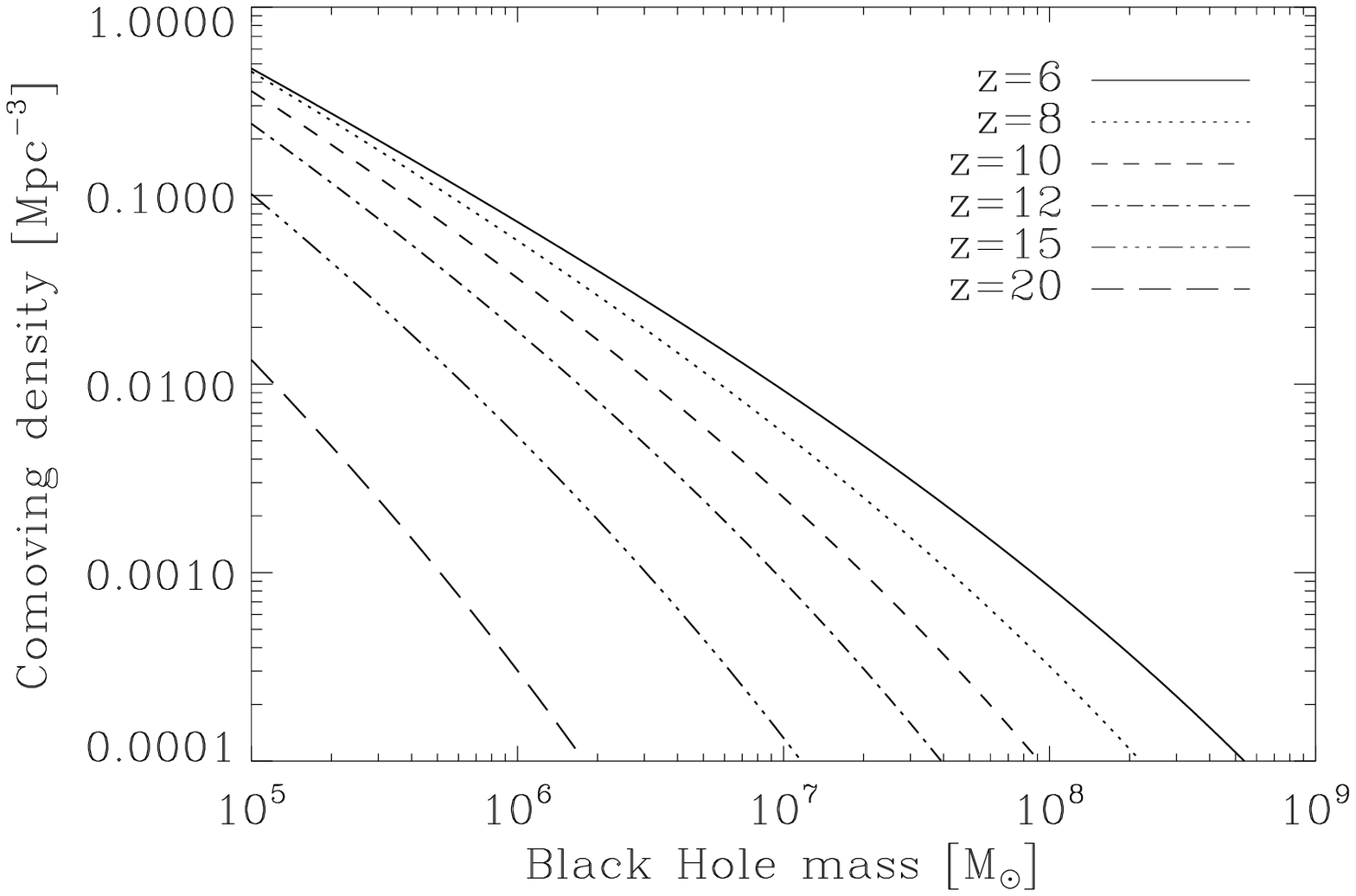}
\caption{The co-moving number density of halos that contain black holes as a function of black hole mass for different redshifts.}
\label{fig:population}
\eef

We model the population of high-redshift black holes in the framework of the Sheth-Tormen formalism \citep{Sheth01}, a modified version of the \citet{Press74} method. Both prescriptions agree reasonably well with simulations of high-redshift structure formation \citep{Jang01, Heitmann06, Reed07, Greif08}. This formalism predicts the comoving number density of halos with different masses, $dn/d\log M_h$, where $M_h$ is the halo mass. We calculate the virial radius $r_{\sm{vir}}$ assuming a mean overdensity of $178$ and estimate the velocity dispersion $\sigma$ as the Kepler velocity at this radius. For simple profiles such as the singular isothermal sphere, the velocity dispersion is independent of position, while in more realistic scenarios, there will be some scatter around this value. We adopt the correlation between black hole mass $M_{\sm{BH}}$ and velocity dispersion $\sigma$ given by \citet{Tremaine02} as
\begin{equation}
\log\left(\frac{M_{\sm{BH}}}{M_\odot} \right) = \alpha+\beta \log\left(\frac{\sigma}{\sigma_0} \right),
\end{equation}
with $\alpha=8.13\pm0.06$, $\beta=4.02\pm0.32$ and $\sigma_0=200$~km/s. The expected black hole population is then given in Fig.~\ref{fig:population}. This shows that a significant number of potential sources is available down to $z\sim12$, which however drops significantly towards higher redshifts.

If black holes accrete efficiently and release the gravitational energy as radiation, the luminosity can be estimated with the Eddington luminosity as
\begin{equation}
L_{\sm{Edd}}=3.3\times10^4 \left(\frac{M_{\sm{BH}}}{M_\odot} \right) L_\odot .
\end{equation}
On the other hand, accretion may also be limited by gas dynamics, and the Bondi solution for spherical symmetric accretion shows that the actual accretion rate may be considerably smaller if the gas is heated by radiative feedback or supernova explosions. As discussed by \citet{Jolley08}, part of the released gravitational energy may go into mechanical feedback like jets and outflows rather than radiation. This is consistent with measurements of the black hole mass - luminosity relation by \citet{Kaspi00}, which indicate that typical luminosities are in fact sub-Eddington. They derive two correlations between black hole mass and luminosity, depending on whether the velocity estimate is obtained taking the mean (case A) or the root mean square~(rms, case B). They use the parametrization
\begin{equation}
M_{\sm{BH}}=a\times10^7\times \left(\frac{\lambda L_\lambda (510\ nm) }{10^{44}\ \mathrm{erg}\ \mathrm{s}^{-1}} \right)^b\ M_\odot.
\end{equation}
In case A, they obtain $a_A=5.71^{+0.46}_{-0.37}$ and $b_A=0.545\pm0.036$, in case B, they find $a_B=5.75^{+0.39}_{-0.36}$ and $b_B=0.402\pm0.034$. Like \citet{Kaspi00}, we adopt a factor of $9$ to correct for the bolometric luminosity. In both cases, the luminosity scales roughly with $M_{\sm{BH}}^2$ and most quasars are sub-Eddington. As shown in the next section, the X-ray background constraints on the black hole population are much less significant with this mass-luminosity relation.

\section{Constraining the relative number of active black holes}\label{constraint}
The population of high-redshift black holes is constrained by the unresolved soft X-ray background (SXRB). The total SXRB in the energy range $0.5-2$~keV is $(7.53\pm0.35)\times10^{-12}$~erg~cm$^{-2}$~s$^{-1}$~deg$^{-2}$, with $94^{+6}_{-7}$ per cent made up of discrete X-ray sources \citep{Moretti03}. The unresolved fraction of the SXRB was re-analyzed by \citet{Dijkstra04}, providing a mean and maximum intensity of the unresolved SXRB flux of $0.35\times10^{-12}$ and $1.23\times10^{-12}$~erg~cm$^{-2}$~s$^{-1}$~deg$^{-2}$, respectively. This corresponds to specific fluxes per energy interval of the order $0.7-2.5$~keV~cm$^{-2}$~s$^{-1}$~sr$^{-1}$~keV$^{-1}$ at $1$~keV, which is the upper limit for possible contributions from the early universe.

\begin{figure}

\includegraphics[scale=0.4]{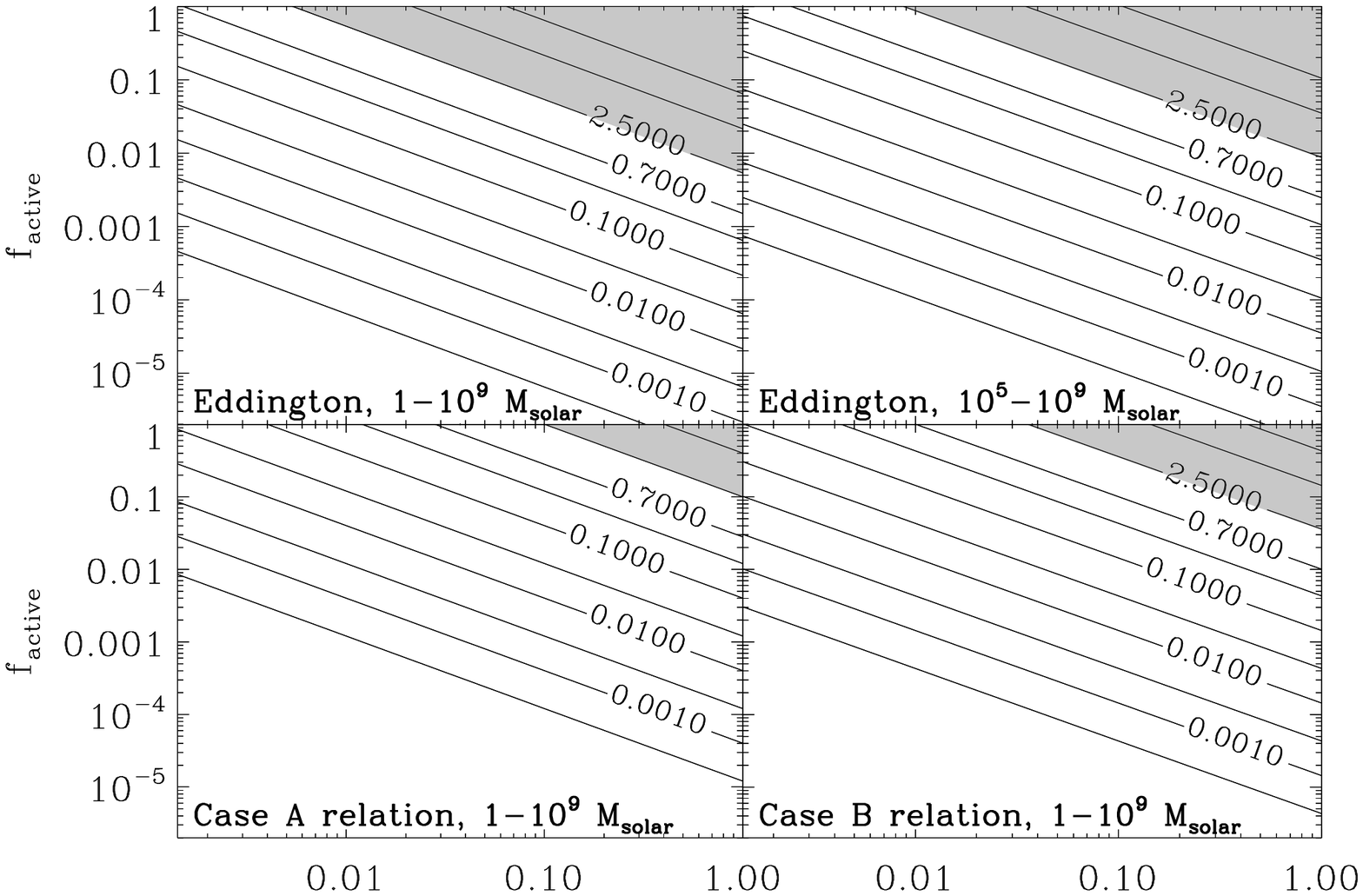}
\caption{The expected contribution to the SXRB in keV~cm$^{-2}$~s$^{-1}$~sr$^{-1}$~keV$^{-1}$ as a function of the active quasar fraction $f_a$ and the spectral parameter $\Phi$, under different assumptions for the typical black hole luminosity. For a spectrum with the same amount of flux in the MCD and the PL component, we expect $\Phi\lesssim0.5$. Upper left: Eddington luminosity, black holes in the mass range from $1$ to $10^9\ M_\odot$. Upper right: Eddington luminosity, black hole masses in the range from $10^5$ to $10^9\ M_\odot$. Lower left: Case A black hole - luminosity relation \citep{Kaspi00}. Lower right: Case B black hole - luminosity relation \citep{Kaspi00}. For the latter models, results are independent of the lower mass cutoff due to the scaling of the employed luminosity relations with mass. In general, we find that models assuming Eddington luminosity find significantly stronger constraints than models following the observed luminosity relation. This implies that the fraction of active quasars may be significantly higher than for Eddington luminosities.    }
\label{fig:constraints}

\end{figure}

\subsection{Assumptions}
For the X-ray sources, we consider spectra of similar type as \citet{Salvaterra05}, consisting of a soft multicolor disk (MCD) component and a power-law (PL) component at high energies \citep{Mitsuda84, Miller03, Miller04}. The MCD component peaks at frequencies
\begin{equation}
h\nu\sim 0.095\ \mathrm{keV} \left(\frac{M_{\sm{BH}}}{10^6\ M_\odot} \right)^{-1/4}
\end{equation}
and the spectral energy distribution (SED) scales with $\nu^{1/3}$ for smaller frequencies and rolls off exponentially for larger frequencies. The MCD component of high-redshift quasars is therefore redshifted to even smaller frequencies and cannot contribute to the unresolved SXRB. For the PL SED, we assume a scaling according to $\nu^{-\alpha}$, with $\alpha\sim1$. It is the PL component of these sources which is redshifted into the SXRB. The spectral range in which these sources emit is not completely clear. Here, we adopt a spectral range from $1$~eV to $500$~keV. The ratio $r$ of the flux in the MCD component to the flux in the PL component is expected to be roughly of order $1$ \citep{Mitsuda84, Miller03}, but we treat it as a free parameter in this analysis. A further fraction $f_s$ of the radiation may be shielded by the inner torus. This fraction is geometry dependent, but may be of the order of $20-30\%$. The amount of luminosity that can potentially contribute to the SXRB thus scales with the effective parameter $\Phi=(1-f_s)/(1+r)$. For typical spectra where the amount of flux in the MCD and the PL component is comparable, we therefore expect $\Phi \lesssim 0.5$. We further introduce the parameter $f_a$, which describes the fraction of active quasars at a given redshift. 

\subsection{Formalism}
\begin{figure}[t]
\includegraphics[scale=0.3]{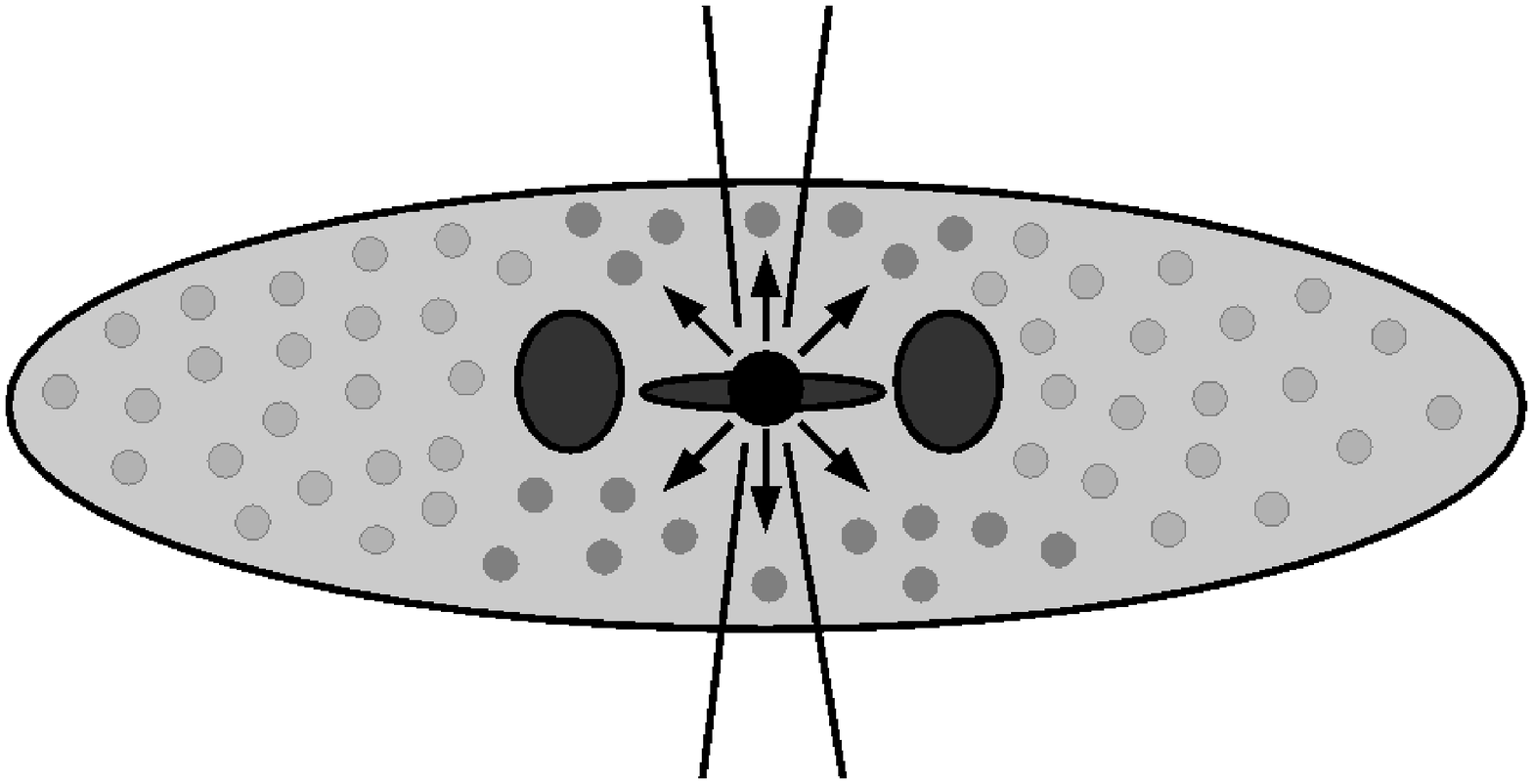}
\caption{An illustration for the inner structure of NGC~1068. It shows the central black hole with the inner accretion disk radiating in X-rays, as well as the inner torus embedded in a large molecular disk consisting of individual high-density clouds. The jet is sketched as well. The torus is expexted to have a size of $\sim3$~pc, while the molecular disk extends out to $\sim100$~pc. In the XDR model considered here, we focus on those clouds that are unshielded by the torus and where emission is stimulated by the impinging X-ray flux.}
\label{fig:illu}
\eef

The contribution to the background intensity is then given from an integration along the line of sight as
\begin{equation}
I_\nu=\frac{c}{4\pi}\int \frac{dz P_\nu ((1+z)\nu),z}{H(z)(1+z)^4},
\end{equation}
where $\nu$ is the observed frequency, $H(z)$ the expansion rate at redshift $z$ and $P_\nu((1+z)\nu,z)$ the mean proper volume emissivity for photons with frequency $\nu(1+z)$ at redshift $z$ (in units of energy per unit time, unit frequency and unit proper volume). To evaluate that, we first calculate the number of photons $N_{\tilde{\nu}}$ of frequency $\tilde{\nu}$ emitted at redshift $z$ within $1$~Mpc$^3$~(comoving), which is given as
\begin{equation}
N_{\sm{tot}, \tilde{\nu}}=\int_{M_{\sm{min}}}^{M_{\sm{max}}} \frac{dn}{d\log M_h} f_a N(\tilde{\nu},M_{\sm{BH}}) d\log M_h,
\end{equation}
where $M_{\sm{min}}$ and $M_{\sm{max}}$ are the (redshift-dependent) minimum and maximum halo masses that yield black hole masses in a reasonable range which we specify below in more detail, and $N(\tilde{\nu},M_{\sm{BH}})$ is the number of photons of frequency $\tilde{\nu}$ emitted per second by a black hole of mass $M_{\sm{BH}}$. The number of baryons $N_b$ in $1$~Mpc$^3$~(comoving) is 
\begin{equation}
N_b=\frac{\rho_b\,1\,\mathrm{Mpc}^3}{1.2 m_p},
\end{equation}
where $\rho_b$ is the comoving baryon density and $m_p$ the proton mass. The emissivity is then 
\begin{equation}
P_\nu=\frac{h\tilde{\nu} N_{\sm{tot}, \tilde{\nu}} }{N_b}\frac{ \rho_b (1+z)^3}{1.2 m_p}.
\end{equation}

\subsection{Constraints}

With the method described above, we calculate constraints for the fraction of active quasars $f_a$ as well as the spectral parameter $\Phi$ for four different scenarios (see Fig. \ref{fig:constraints}). The first scenario assumes that black holes populate the mass range from $1$ to $10^9\ M_\odot$ and radiate at Eddington luminosity. This may be the case if a sufficient number of black holes is provided as remnants from the first stars. In this case, one finds rather tight constraints that do not allow high values for $f_a$ and $\Phi$ simultaneously. 

However, as the initial mass function (IMF) in the early universe is not yet understood, it is unclear whether such stellar mass black holes are present in large numbers. As discussed in the introduction, supermassive black holes may in fact have more massive progenitors that form through direct gas collapse in larger halos. We assume here a lower cutoff for the black hole mass of $\sim10^5\ M_\odot$. That allows for an increase in the fraction of active quasars by a factor of a few. 

Finally, as discussed in \S \ref{population}, the Eddington luminosity is only an upper limit to the actual black hole luminosity. It depends on the gas supply as well as the amount of energy going into mechanical feedback. To address alternative scenarios as well, we calculate the constraints assuming the black hole mass - luminosity relations given by \citet{Kaspi00}. As a consequence, a higher fraction of active quasars is consistent with the observed background for parameters $\Phi$ between $0.1-1$. Due to the steepness of the luminosity function, the results are independent of the adopted lower cutoff for the black hole mass. In case A, $f_a\sim50\%$ seems feasible for $\Phi\sim0.2$, while in case B, $f_a\sim50\%$ requires $\Phi\sim0.1$. For a spectrum with equal amount of flux in the MCD and the PL component ($\Phi\lesssim0.5$), active quasar fractions of $10-30\%$ can be adopted.

With respect to the observational search for the first quasars at high redshift, the two latter scenarios are particularly favorable, as they are consistent with a considerably larger number of active quasars in the early universe. The main caveat here is the question whether the observed relation between black hole mass and luminosity can indeed be used at these early times. So far quasars have always appeared as highly evolved systems at any redshift observed, which may imply that this relation indeed holds. In any case, the first scenario assuming Eddington luminosity provides a firm lower limit of about $1\%$ for the fraction of active quasars if $\Phi$ is assumed to be of order $1$.

\section{NGC~1068: A case study}\label{ngc1068}

\begin{figure}[p]
\includegraphics[scale=0.55]{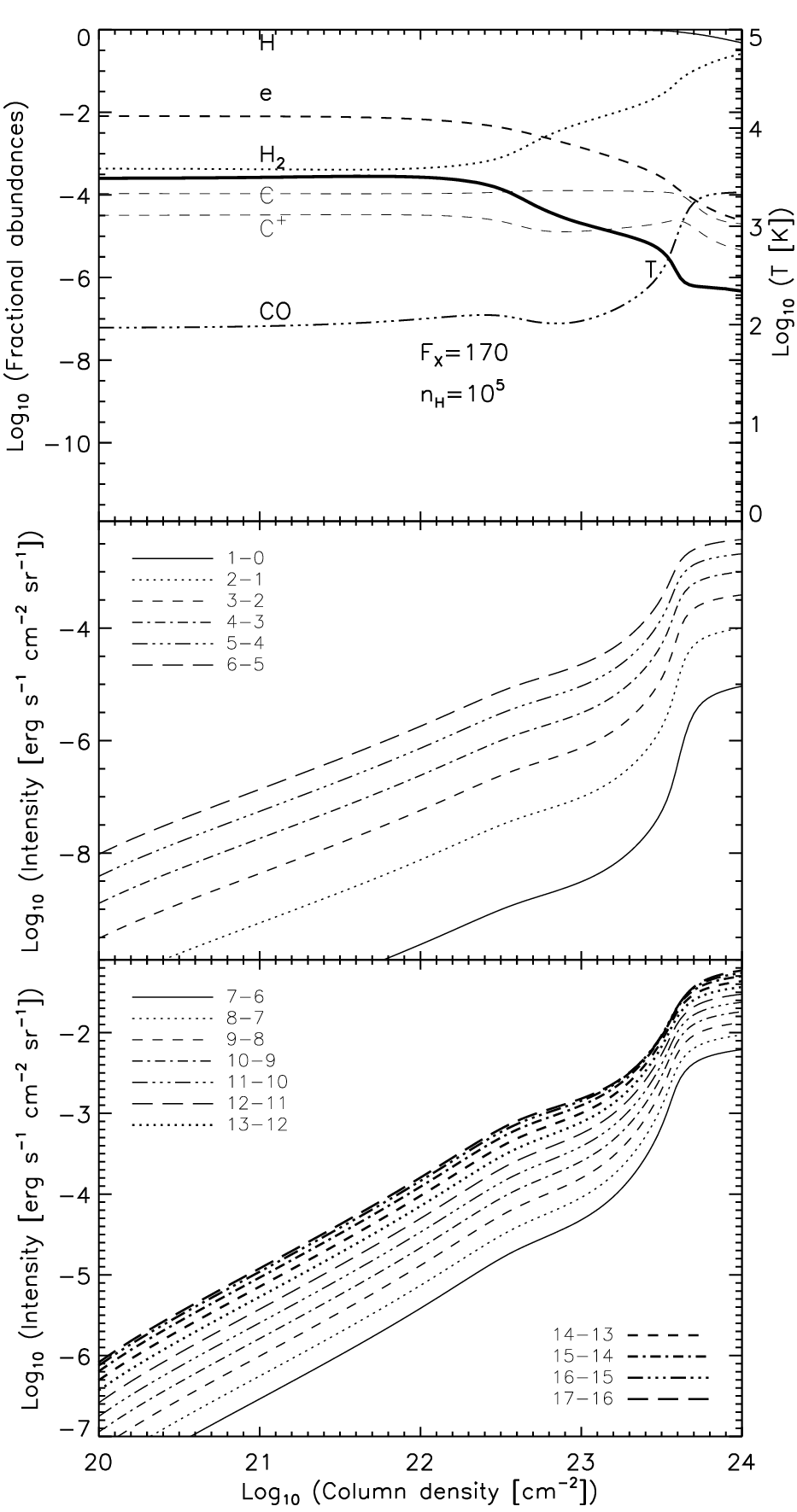}
\caption{A model for the X-ray chemistry in NGC~1068. The adopted flux impinging on the cloud is $170$~erg~s$^{-1}$~cm$^{-2}$. The adopted density is $10^5\ $cm$^{-3}$. Top: The abundances of different species as a function of column density. Middle: The low-$J$~\COI lines as a function of column density. Bottom: The high-$J$~\COI lines as a function of column density. }
\label{fig:ngc1068}
\eef
In this section, we review known observational properties of NGC~1068, derive predictions that will be relevant for high-redshift observations with ALMA and JWST and generalize them to objects with different X-ray fluxes.

\subsection{Observed properties}\label{observed}

As discussed in the Introduction, the nearby AGN NGC~1068, a Seyfert~$2$~galaxy at a distance of $14.4$~Mpc, has been studied in great detail and with high resolution. As the wealth of available data could easily give rise to a review of its own, we restrict the discussion here to only those properties that will be relevant with respect to observations at high-redshift with ALMA and JWST. 

The mass of the central black hole has been estimated from VLBI water maser observations to be $\sim10^7\ M_\odot$ with a luminosity of $50\%$ of the Eddington limit \citep{Greenhill96}. Even though the X-ray emission from the central black hole is shielded by an absorber on the line of sight, X-rays that are scattered into the line of sight have been detected unambiguously \citep{Pounds06}. Observations of rotational \COI lines and the \HzI~$2.12\ \mathrm{\mu}$m line indicated an inner structure consisting of an extended molecular disk and a toroidal structure which completely absorbs the X-ray emission along the line of sight. It is slightly tilted with respect to the disk \citep{Galliano03}. The \HzI emission map of \citet{Galliano02} shows a bright knot $70$~pc in the east of the central engine, with a size of $\sim35$~pc. The knot is also visible in \COI \citep{Galliano03}. As they show, the observed fluxes can be explained with the X-ray excitation model of \citet{Maloney96} when the following parametrization is adopted:
\begin{itemize}
\item The central engine is a power-law X-ray source with spectral slope $\alpha=-0.7$ and luminosity of $10^{44}$~erg~s$^{-1}$ in the $1-100$~keV range.
\item We observe emission from molecular clouds with a density of $10^5$~cm$^{-3}$, a column of $10^{22}$~cm$^{-2}$ at a distance of $70$~pc and solar metallicity.
\end{itemize}
This region of bright emission consists therefore of a number of small clouds with densities of $10^5$~cm$^{-3}$, which in total cover a significant part of the telescope beam. In addition, there may be further contributions from gas at lower densities, which emits less efficiently, but fills a larger volume. For the X-ray luminosity of NGC~1068, one cannot expect \COI line emission much closer to the black hole, as the higher X-ray flux then would destroy \COI very effectively, unless it is shielded by some absorber. The black hole luminosity thus regulates the distance to the central source at which the \COI emission peaks. This situation is sketched in Fig.~\ref{fig:illu}.

Further studies also measured the fluxes from other species, including the relevant fine-structure lines. We refer here in particular to the data of \citet{Spinoglio05}, which are used in \S \ref{ALMA} to calculate the expected fluxes from higher redshift. The situation is similar for the observed amount of continuum emission. Here we use the data of \citet{Brauher08}. 

As JWST will see the stellar light from high-redshift galaxies, the star formation rate is a highly relevant quantity for estimates regarding the expected flux. Indeed, observations of NGC~1068 find a starburst component from a circumnuclear ring of $\sim3$~kpc in size, with a stellar mass of $\sim10^6\ M_\odot$ and an age of $5$~Myr \citep{Spinoglio05}. On scales of a few hundred parsecs, one finds star formation rates of a few times $10\ M_\odot$~yr$^{-1}$~kpc$^{-2}$, which can reach $\sim100\ M_\odot$~yr$^{-1}$~kpc$^{-2}$ in the inner few ten parsecs \citep{Davies07}. This is close to the star formation rate in Eddington-limited starbursts as suggested by \citet{Thompson05}. The total star formation rate of NGC~1068 therefore is $\sim10\ M_\odot\ $yr$^{-1}$.

\subsection{Predicted properties}\label{predicted}
High redshift and high spatial resolution ($\sim 0.01''$) observations with ALMA will allow us to observe redshifted high-$J$ CO lines, $J>9$ (see \S \ref{ALMA} for details). For local sources, these lines are difficult to detect from the ground. They provide important diagnostics for the power and impact of the accreting black hole \citep{Spaans08}. We use the X-ray dominated region (XDR) code of \citet{Meijerink05}, an improved version of the above mentioned XDR-model of \citet{Maloney96}, and solve the 1D-radiation transport equation for the \COI fluxes through the inferred cloud complex. This code has been used to calculate detailed diagnostics for XDR observations \citep{Meijerink07} and was successfully applied to infrared and (sub-)millimeter observations of luminous infrared galaxies \citep{Loenen08}.

\begin{figure}[p]
\includegraphics[scale=0.55]{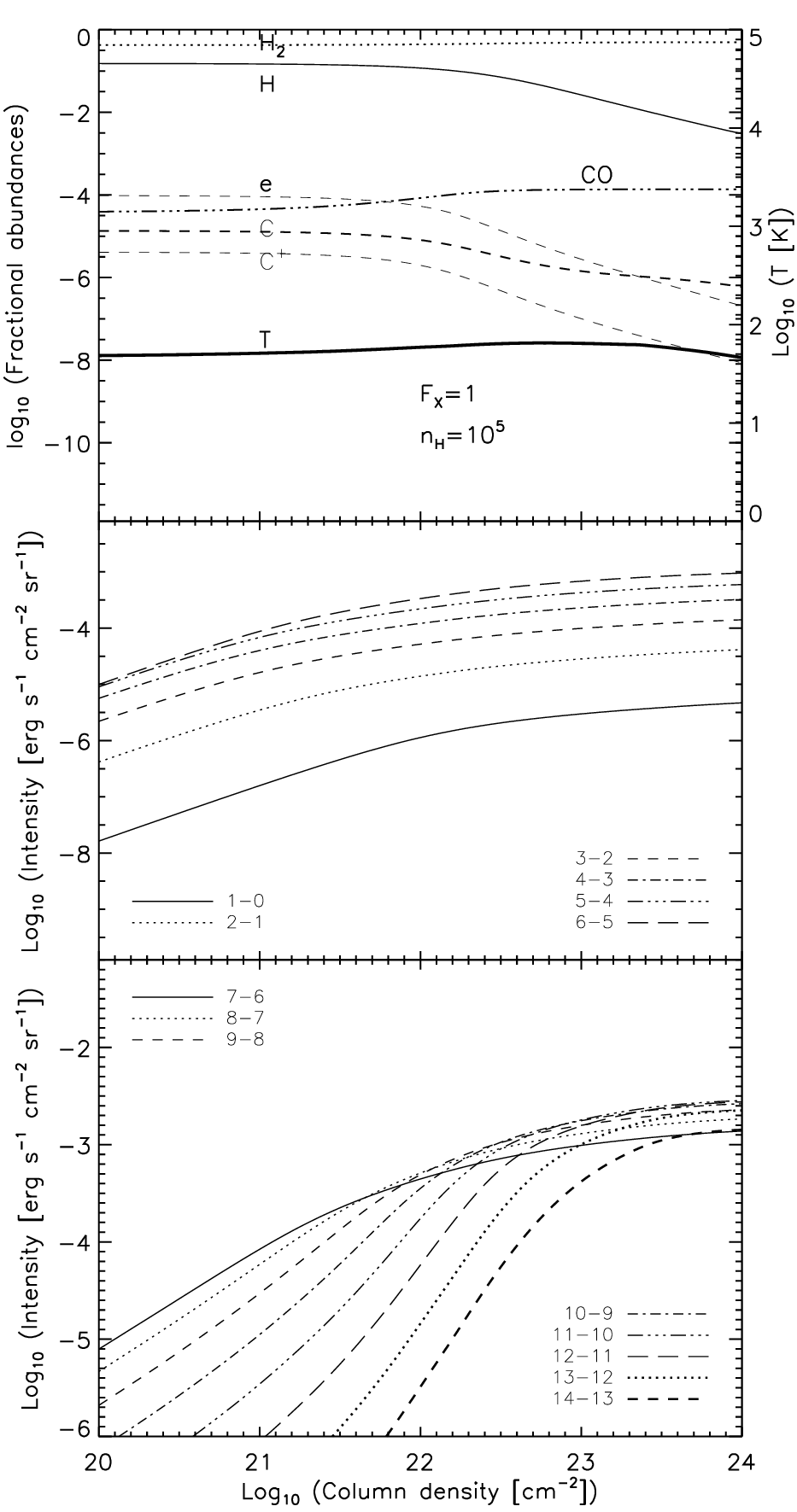}
\caption{The X-ray chemistry in a system with X-ray flux of $1$~erg~s$^{-1}$~cm$^{-2}$ impinging on the cloud. The adopted density is $10^5\ $cm$^{-3}$. Top: The abundances of different species as a function of column density. Middle: The low-$J$~\COI lines as a function of column density. Bottom: The high-$J$~\COI lines as a function of column density. }
\label{fig:ngc1}
\eef

\begin{figure}[p]
\includegraphics[scale=0.55]{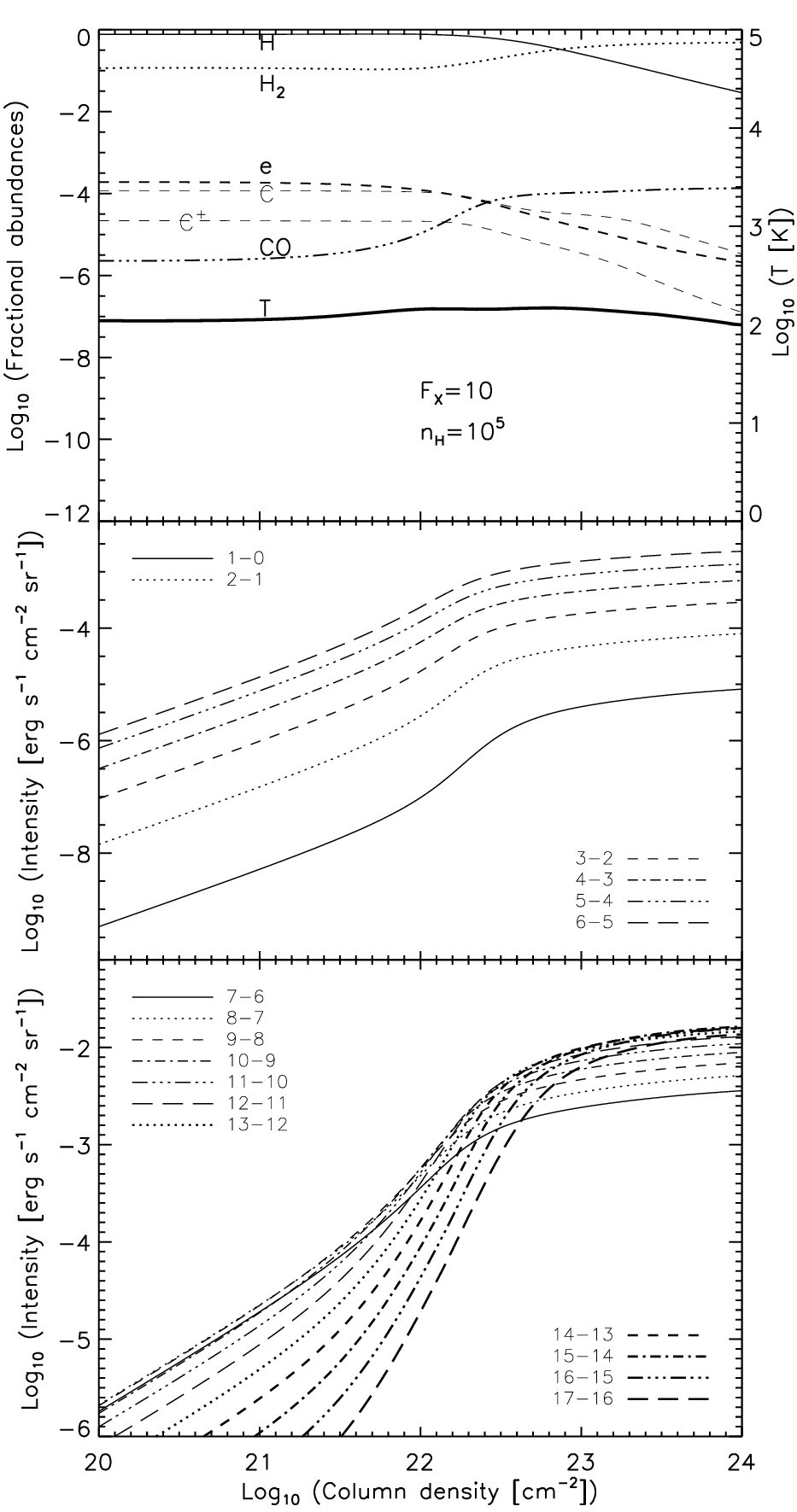}
\caption{The X-ray chemistry in a system with X-ray flux of $10$~erg~s$^{-1}$~cm$^{-2}$ impinging on the cloud. The adopted density is $10^5\ $cm$^{-3}$. Top: The abundances of different species as a function of column density. Middle: The low-$J$~\COI lines as a function of column density. Bottom: The high-$J$~\COI lines as a function of column density. }
\label{fig:ngc10}
\eef

We first calculate the chemical abundances and the expected \COI line intensities for the model used by \citet{Galliano03} as discussed in \S~\ref{observed}. While they adopted a column density of $10^{22}\ \mathrm{cm}^{-2}$, we are extending the calculation to larger columns of $10^{24}\ \mathrm{cm}^{-2}$, corresponding to the typical column needed to absorb $5$~keV photons. This allows us to generalize these results to systems with larger column densities and Compton thick material. The results of the calculations are shown in Fig.~\ref{fig:ngc1068}. We note that the fiducial gas density of $10^5$~cm$^{-3}$ has little impact on our results, unless it drops to below $10^{4.5}$~cm$^{-3}$. Above this limit, the XDR properties are determined by the ratio of X-ray flux to gas density. In addition, we have checked that even if the entire disk of $\sim100$~pc was at low densities of $10^3$~cm$^{-3}$, the high-$J$~\COI line intensities would still be $\sim10^{-5}$~erg~s$^{-1}$~cm$^{-2}$~sr$^{-1}$, as the low-density gas extends over much larger spatial scales and thus reaches larger or comparable column densities.

The strong X-ray flux of $\sim170$~erg~s$^{-1}$~cm$^{-2}$ in NGC~1068 suffices to make the gas essentially atomic and leads to high temperatures of $\sim3000$~K, as well as relatively low \COI abundances of the order $10^{-7}$. However, the \COI intensity is high, due to the strong excitation in the hot gas. For a column of $10^{22}\ \mathrm{cm}^{-2}$, our results appear of the same magnitude as in the model of \citet{Galliano03}. For larger columns, the temperature gradually decreases, the gas becomes molecular and \COI gets more abundant, and we find intensities of the order $10^{-2}$~erg~s$^{-1}$~cm$^{-2}$~sr$^{-1}$ in the high-$J$~\COI lines.

We need to account for the fact that many systems may accrete considerably below the Eddington limit as discussed above. Also, we want to consider quasars with somewhat smaller black hole masses of the order $10^6\ M_\odot$, as their number density is larger by an order of magnitude (see Fig.~\ref{fig:population}). Therefore, we calculate models with less X-ray flux as well. An extreme case with $\sim1$~erg~s$^{-1}$~cm$^{-2}$ is shown in Fig.~\ref{fig:ngc1}. In this model, we find lower temperatures of $\sim70$~K, a large fraction of molecular gas and \COI abundances of the order $10^{-4}$. While the lower temperature tends to decrease the \COI line intensities, they are still enhanced due to the larger \COI abundance. Above a column of $10^{23}\ \mathrm{cm}^{-2}$, the intensities increase rather slowly as the lines become optically thick. 

As an intermediate scenario, we consider an X-ray flux of $\sim10$~erg~s$^{-1}$~cm$^{-2}$ (see Fig.~\ref{fig:ngc10}). In this model, the temperature is increased to $\sim100$~K. The \COI abundance is initially of the order $3\times10^{-6}$ and increases to $\sim10^{-4}$ for larger columns. For columns less than $10^{22}\ \mathrm{cm}^{-2}$, the intensities are thus reduced by about an order of magnitude compared to the previous case, while they are increased by an order of magnitude for larger columns. 

From an observational perspective, it is highly interesting that even X-ray fluxes of $\sim1$~erg~s$^{-1}$~cm$^{-3}$ lead to significant \COI line intensities. At a column of $10^{22}\ \mathrm{cm}^{-2}$, the high-$J$~\COI line intensity in all models was of the order $10^{-4}-10^{-3}$~erg~s$^{-1}$~cm$^{-2}$~sr$^{-1}$, and it may increase up to values of $10^{-3}-10^{-1}$~erg~s$^{-1}$~cm$^{-2}$~sr$^{-1}$ for columns of the order $10^{24}\ \mathrm{cm}^{-2}$.

\section{High-redshift observations with ALMA and JWST}\label{observables}

\begin{table}[htdp]
\begin{center}
\begin{tabular}{ccccc}
freq. [GHz]  & $\theta{\sm{res}}$~[''] & $S_l$~[mJy] & $S_c$~[mJy] & $\theta_{\sm{beam}}$~[''] \\
\hline
 $84-116$  & $0.034$ & $8.9$ & $0.06$ & $56$ \\
 $125-169$ & $0.023$ & $9.1$ & $0.07$& $48$ \\
 $163-211$ & $0.018$ & $150$ & $1.3$& $35$ \\
 $211-275$ & $0.014$ & $13$ & $0.14$& $27$ \\
 $275-373$ & $0.011$ & $21$ & $0.25$& $18$ \\
 $385-500$ & $0.008$ & $63$ & $0.86$& $12$ \\
 $602-720$ & $0.005$ & $80$ & $1.3$& $9$ \\
\hline
\end{tabular}
\end{center}
\caption{Frequency range, angular resolution $\theta_{\sm{res}}$, line sensitivity $S_l$ and continuum sensitivity $S_c$ in mJy for one hour of integration time and primary beam size $\theta_{\sm{beam}}$ for bands $3-9$ of ALMA. 3 more bands might be added in the future, band $1$ around $40$~GHz, band $2$ around $80$~GHz and band $10$ around $920$~GHz, which will have similar properties as the neighbouring bands.}
\label{tab:ALMA}
\end{table}%

Future observations with ALMA and JWST provide a new opportunity to find the first quasars at high redshift. Their unprecedented sensitivity permits observations of less luminous sources and at higher redshift and we can therefore find quasars with black hole masses of $10^6-10^7\ M_\odot$, which may be the progenitors of the known sources with $\sim10^9\ M_\odot$ black holes. In addition, their high spatial and spectral resolution can provide detailed information about the host galaxies of those bright high-redshift quasars that we already know. In this section, we predict the main observables for ALMA and JWST, provide estimates for the number of high-redshift sources and discuss also the importance of follow-up observations on known high-redshift quasars.

\subsection{Observables for ALMA}\label{ALMA}
In the frequency range of ALMA, it is possible to observe redshifted emission from high-$J$~\COI lines and many fine-structure lines. The main properties of ALMA are summarized in Table~\ref{tab:ALMA}. For one hour of integration time, ALMA can detect line fluxes of $\sim10$~mJy at $3\sigma$. This is particularly interesting with respect to the high-$J$~\COI intensities calculated in \S~\ref{predicted}. At redshift $z=8$, a resolution of $\sim0.01$'' corresponds to a physical scale of $\sim49$~pc. A source of such size would fill a solid angle of $\sim7\times10^{-15}$~sr. With a fiducial velocity dispersion of $20$~km/s and a line intensity of $10^{-3}$~erg~s$^{-1}$~cm$^{-2}$~sr$^{-1}$, this corresponds to a flux of $\sim30$~mJy, which is therefore detectable in less than one hour. In the calculations above, we found lower intensities only for models with very high X-ray fluxes and small columns (see Fig.'s~\ref{fig:ngc1068}~to~\ref{fig:ngc10}). Even in this case, we still expect a significant flux of $\sim3$~mJy. 

\begin{table}[htdp]
\begin{center}
\begin{tabular}{llll}
Observable & $\lambda$~[$\mathrm{\mu}$m] & $\varphi~[mJy]$ & Redshift \\
\hline
high-$J$ \COI lines  &$\sim300$ &  $3-30$   &  $z>5$ \\
$[$O I$]$~$\ ^3P_1\rightarrow^3P_2$& $63.2$ & $\sim20$   &  $z>5.7$ \\
$[$O~III$]$~$^3P_1\rightarrow^3P_0$& $51.8$ & $\sim20$  &  $z>3.8$ \\
$[$N~II$]$~$^3P_2\rightarrow^3P_1$ & $121.9$ &  $\sim7.6$  &  $z>2.5$ \\
$[$O I$]$~$\ ^3P_0\rightarrow^3P_1$ & $145.5$  & $\sim3.5$   &  $z>1.9$ \\
$[$C II$]$~$\ ^2P_{3/2}\rightarrow^2P_{1/2}$  & $157.7$ & $\sim69$   &  $z>1.7$ \\
$[$S III$]$~$\ ^3P_1\rightarrow^3P_0$ & $33.5$ & $\sim1.5$   &  $z>11.5$ \\
$[$Si II$]$~$\ ^2P_{3/2}\rightarrow^2P_{1/2}$ & $34.8$ & $\sim2.6$   &  $z>11.1$ \\
dust continuum  & $\sim500$ & $\sim0.05$ &  $z>0$  \\
\hline
\end{tabular}
\end{center}
\caption{The main observables for ALMA, as discussed in \S~\ref{ALMA}. The estimated flux $\varphi$ for the high-$J$~\COI lines assumes that one typical resolution element of $\sim0.01''$ is comparable to the source size, while the other fluxes correspond to the fluxes observed for NGC~1068 if it were placed at $z\sim8$ (for lines only observable at $z>10$, we placed it at $z=12$). The rest frame wavelength is given as $\lambda$. The line and continuum sensitivities are given in Table \ref{tab:ALMA}.}
\label{tab:observables}
\end{table}%

 As discussed by \citet{Finkelstein08}, there are further contributions from dust emission which may be relevant for ALMA as well, because the continuum sensitivity is higher by two orders of magnitude. For NGC~1068, \citet{Brauher08} report a flux of $85.04$~Jy at $25$~$\mathrm{\mu}$m and $224$~Jy at $100$~$\mathrm{\mu}$m. For $z>5$, such fluxes would be redshifted into the ALMA bands. An interpolation to $500$~$\mathrm{\mu}$m yields a flux of $\sim700$~Jy, which would be redshifted into ALMA band 3 for $z\sim5$. For the same emission at this redshift, we would therefore expect a flux of $0.06$~mJy, which matches with the band sensitivity. For objects at $z=8$, we need to consider emission at $\sim330$~$\mathrm{\mu}$m which is redshifted into band 3, where the extrapolation yields $\sim520$~Jy. This yields an expected flux of $\sim0.05$~mJy, which is detectable within a few hours of integration time. Such emission may be even stronger if the dust is heated further by ongoing starbursts.


In addition to the high-$J$~\COI lines, it seems likely that other lines can be detected with ALMA as well. For our fiducial AGN NGC~1068, \citet{Spinoglio05} measured a number of different fine-structure lines with the Long Wavelength Spectrometer (LWS) on board the Infrared Space Observatory (ISO). From their sample, we have identified several lines that are interesting for observations at high redshift. Assuming the same source at larger redshift, we find typical fluxes that could be detected with ALMA within one to several hours (see Table~\ref{tab:observables}). The high angular resolution of ALMA may even allow to resolve and detect the higher fluxes from the inner core. The detection of the \OI and \CII fine-structure lines will be particularly interesting, as the sum of the [O~I]~$63$~$\mathrm{\mu}$m, the [C~II]~$158$~$\mathrm{\mu}$m and the \COI lines is close to the total cooling rate of the gas, given that above $10^{-2}$ solar metallicity the cooling via \HzI rotational lines is only a small contribution. 

Since at high densities in XDRs the thermal balance establishes rapidly, the total cooling flux equals the heating flux. This allows one to infer the impinging X-ray flux, while the sum of the fine-structure and CO lines divided by the heating efficiency is always a measure of the X-ray flux that has been transformed into heat. The heating efficiency of $\sim10\%$ is largely independent of energy and the spectral shape of the flux \citep{Dalgarno99}. 
For a spatially resolved source, the linear size of the cooling gas can be obtained from the spectroscopic redshift. The corresponding X-ray flux on that size scale can then be converted into a luminosity. The observed profiles of the line width may be used to derive further estimates for the black hole mass and the accretion rate \citep{Kaspi00, McLure02, Vestergaard02, Warner03, Dietrich04}. In addition, a detection of multiple \COI lines provides the density and column density distribution as a function of distance from the central engine.

Of course, the metallicity is also a caveat regarding some of the observables discussed here. While this work assumes that it is close to the solar value, the situation for low-metallicity gas has been explored previously by \citet{Spaans08}. For metallicities $Z>10^{-2}\ Z_\odot$, they find intensities in the range $10^{-6}-10^{-3}$~erg~s$^{-1}$~cm$^{-2}$~sr$^{-1}$. For even lower metallicities, the \COI lines are hard to detect, but emission from \HeHII and H$_3^+$ becomes more prominent. Particularly interesting at very low metallicities is also the strong emission in \HzId, which will be redshifted into ALMA band 10 for sources at $z>10$ if this band is realized. While such low metallicities seem unlikely at $z\sim8$, it may become more relevant at these higher redshift.

The choice of metallicity may be guided by other observables like the amount of dust emission, the detection of a strong starburst by JWST as well as theoretial considerations regarding the observability of \COI~lines in low-metallicity gas \citep{Spaans08}. Such estimates may be refined further employing follow-up observations with other telescopes that search for the UV N$^{4+}$ and C$^{3+}$ lines. Their line ratio provides an important diagnostic for the metallicity and has been used to derive the metallicity in the known high-redshift quasars \citep{Pentericci02}.

\subsection{Observables for JWST}\label{JWST}

In the frequency range of JWST, it is possible to detect redshifted stellar light, the AGN continuum and dust. While such detections will not allow to identify the object unambigously as a quasar, they could be used to provide a source list for more detailed follow-up observations with ALMA, which would allow us to do this identification through the shape of the \COI line SED \citep{Spaans08}. In addition, they provide complementary information about star formation in these systems. Quasars with unobscured broad-line regions may also be visible in further lines. In this section, we will focus on the Near Infrared Camera (NIRCam), which will provide high sensitivity of the order a few nJy for a $3\sigma$ detection at $0.7-4.44\ \mathrm{\mu}$m wavelengths after an integration time of $10^4$~s (see Table~\ref{tab:JWST}) and a field of view of $(2.16')^2$. It will consist of $4096^2$ pixels, with an angular resolution of $0.0317''$ per pixel for the short wavelengths and $0.0648''$ per pixel for the long wavelengths.

\begin{figure}[t]
\includegraphics[scale=0.55]{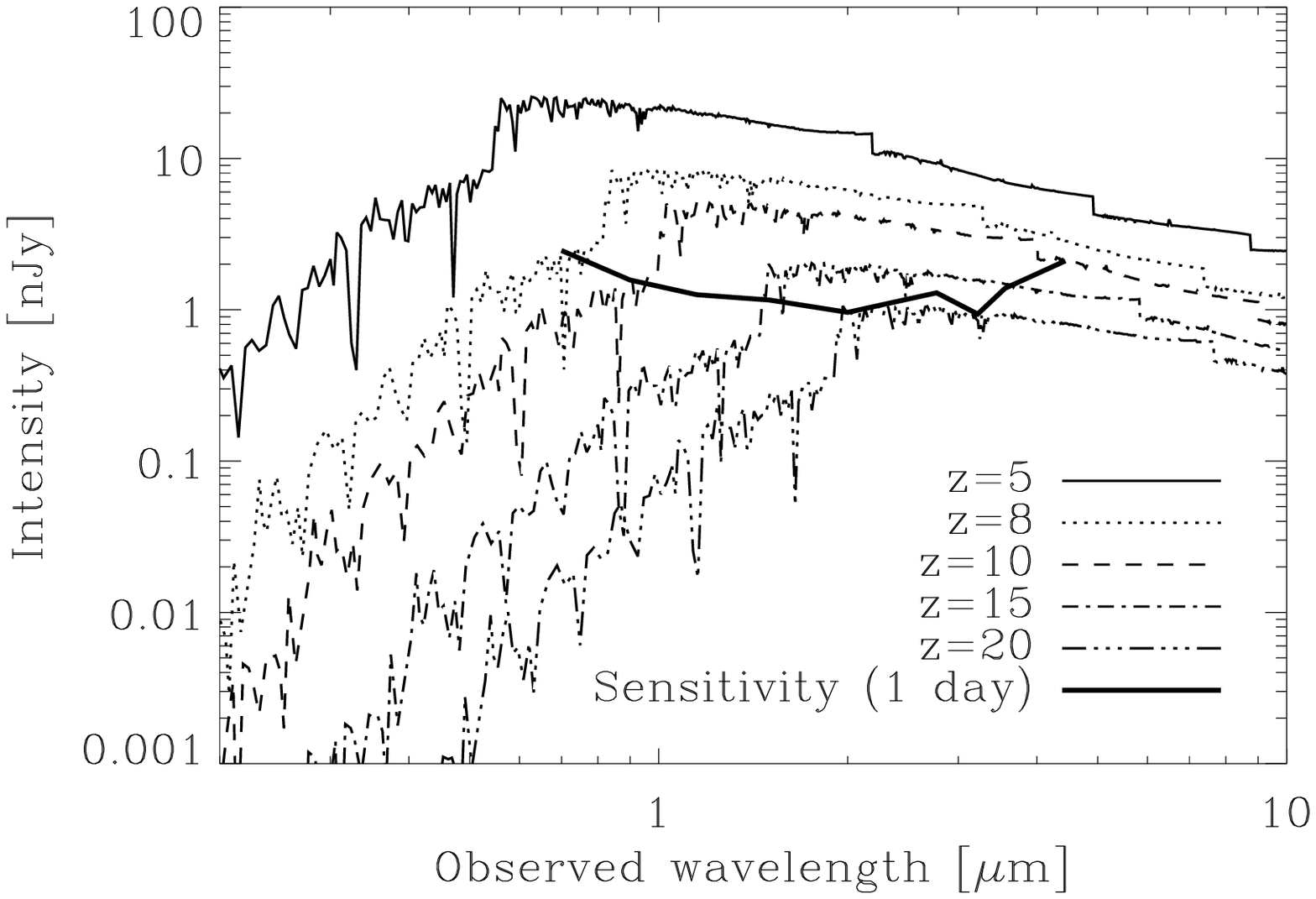}
\caption{The observed spectrum of a starburst with $10\ M_\odot$~yr$^{-1}$ with an age of $5$~Myr located at different redshifts. This is compared to the NIRCam sensitivities for $3\sigma$ detection after an integration time of one day. For $z\lesssim10$, such starbursts can be detected within a few hours.}
\label{fig:starlight}
\eef

As discussed in \S~\ref{observed}, we expect a star formation rate of $\sim10\ M_\odot$~yr$^{-1}$ in an object like NGC~1068. To calculate the corresponding spectrum, we employ the evolutionary synthesis code Starburst99 \footnote{http://www.stsci.edu/science/starburst99/} \citep{Leitherer99, Vazquez05}. We adopt their standard IMF with an exponent of $1.3$ in the range $0.1-0.5\ M_\odot$, and $2.3$ in the range $0.5-100\ M_\odot$  and take the spectrum after a simulation time of $5$~Myr, which corresponds to the age of the circumnuclear starburst ring in this object. We then calculate the expected flux for this starburst at different redshifts. The results are shown in Fig.~\ref{fig:starlight} and compared to the NIRCam sensitivity. For an integration time of one day, we find that such a starburst can be detected with $3\sigma$ at redshift $z=15$. For redshifts $z\lesssim10$, even a few hours are sufficient for detection.

For comparison, we also consider a model with a star formation rate of $\sim1\ M_\odot$~yr$^{-1}$ (see Fig.~\ref{fig:starlight2}). In this case, an integration time of one day is still sufficient to find a starburst at $z\sim5$. A more extended integration time of $10$ days even suffices to find such starbursts down to $z\lesssim10$. This is still rather short compared to proposed surveys searching for light from the first galaxies, which require integration times of $0.4$~years \citep{Windhorst06}.

For the AGN continuum, \citet{Spinoglio05} present a model that peaks at $0.1-0.3$~eV with a flux of $\sim0.01$~Jy. Such radiation will be redshifted into the frequency range of JWST for $z>5$. We would expect a flux of $\sim0.9$~nJy at $z=5$ and $\sim0.3$~nJy at $z=8$. The first would be detectable after an integration time of about three days, and the latter within a time of about $15$~days. In practice, this radiation component will of course overlap with the radiation from the starburst and just leads to a mild increase in the total radiation budget.

\begin{table}[htdp]
\begin{center}
\begin{tabular}{ll}
Wavelength  & Sensitivity \\
\hline
$0.70$ $\mathrm{\mu}$m  &  $7.24$ nJy  \\
$0.90$ $\mathrm{\mu}$m &  $4.62$ nJy  \\
$1.15$ $\mathrm{\mu}$m &  $3.70$ nJy   \\
$1.50$  $\mathrm{\mu}$m &  $3.41$ nJy  \\
$2.00$ $\mathrm{\mu}$m  & $2.83$ nJy  \\
$2.77$ $\mathrm{\mu}$m& $3.81$ nJy  \\
$3.22$ $\mathrm{\mu}$m& $2.75$ nJy \\
$3.56$ $\mathrm{\mu}$m & $4.08$ nJy  \\
$4.44$ $\mathrm{\mu}$m &  $6.18$ nJy \\
\hline
\end{tabular}
\end{center}
\caption{The NIRCam sensitivity of JWST for a detection at $3\sigma$ after an integration time of $10^4$~s. These values assume that the wide filter is used, which is appropriate for continuum fluxes as considered here.}
\label{tab:JWST}
\end{table}%

\begin{figure}[t]
\includegraphics[scale=0.55]{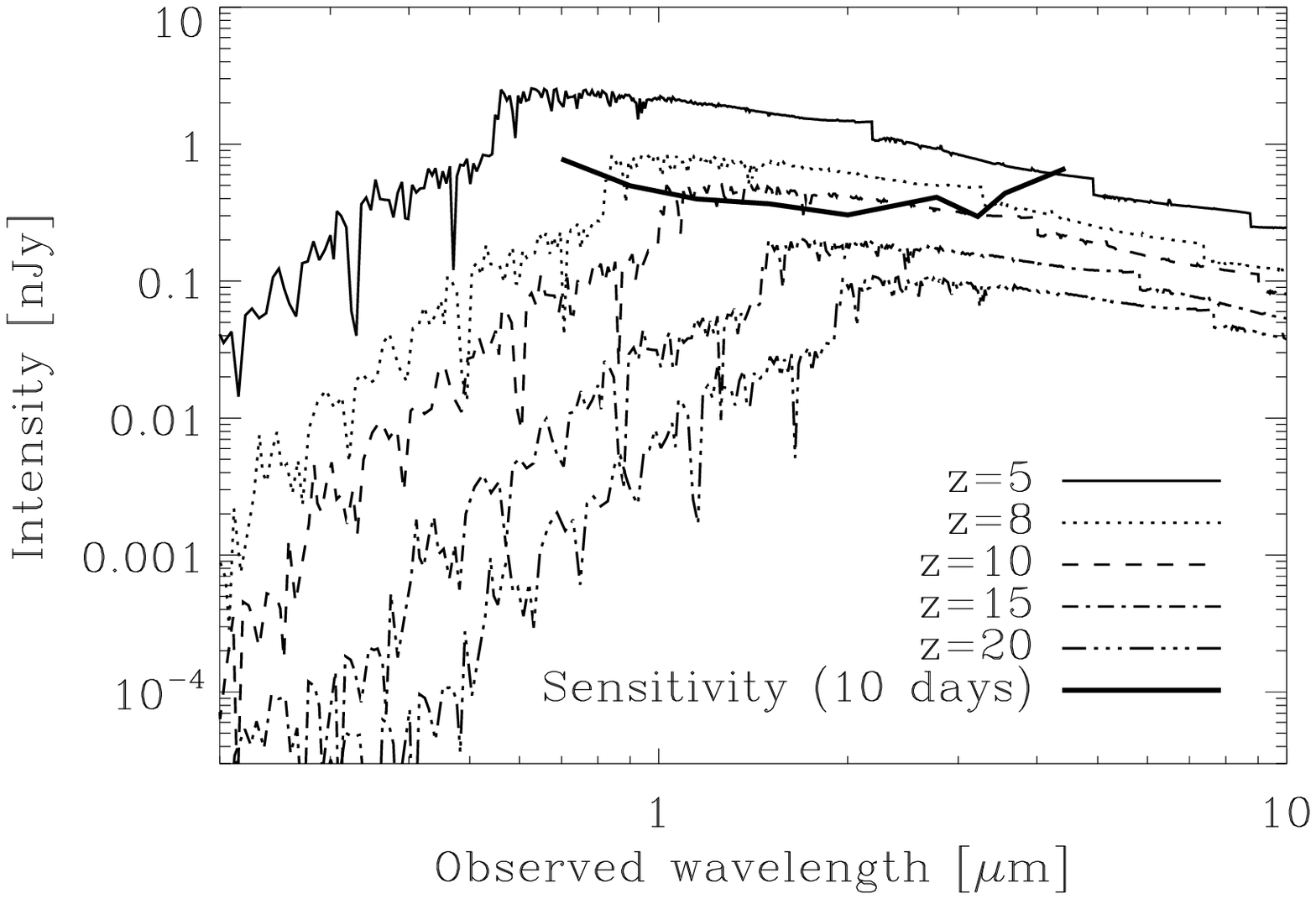}
\caption{The observed spectrum of a starburst with $1\ M_\odot$~yr$^{-1}$ with an age of $5$~Myr located at different redshifts. This is compared to the NIRCam sensitivities for $3\sigma$ detection after an integration time of $10$ days. This allows $3\sigma$ detection of such starbursts for $z\lesssim10$.}
\label{fig:starlight2}
\eef

Additional contributions are also present from redshifted Lyman $\alpha$ radiation, which are particularly enhanced in the presence of a starburst or a supermassive black hole. For example in a halo containing $M_h=10^{11}\ M_\odot$, \citet{Dijkstra06a, Dijkstra06b} find Lyman~$\alpha$ luminosities of $\sim10^{42}$~erg~s$^{-1}$. Considering such a source at high redshift and spreading this energy over a frequency range of $\sim5\times10^4$~GHz, corresponding to the width of the detector, yields a flux of $\sim8$~nJy at $z=5$, and $\sim2$~nJy at $z=8$. As they find the luminosity to scale with $M_h^{5/3}$, this contribution becomes less significant in smaller systems. Such a flux could be detected within several hours.

If in addition the Mid Infrared Instrument (MIRI) camera is employed, which has a field of $1.88'\times1.27'$ and observes at $5-28\mu$m with sensitivities ranging from $60-8000$~nJy in $10^4$~s for a $3\sigma$ detection and with $R\sim3000$ an intrinsic line width of $\sim100$~km~s$^{-1}$, one may aim at detecting the H~$\alpha$ line at $656.3$~nm. This would allow us to derive black hole masses and accretion rates from the line width, and is complementary to to a derivation of the X-ray luminosity with ALMA \citep{Kaspi00, McLure02, Vestergaard02, Warner03, Dietrich04}. \citet{Kaspi00} give light curves of the H~$\alpha$ and H~$\beta$ fluxes for a sample of quasars. We have estimated the fluxes obtained from the first five sources in their sample, assuming these objects would be shifted to redshift $z\sim8$. The H~$\alpha$ line would then be redshifted to $\sim5.9$~$\mathrm{\mu}$m, where the MIRI camera has a sensitivity of $60$~nJy for $3\sigma$ detection in $10^4$~s. As we find H~$\alpha$ fluxes between $100-700$~nJy, a detection is possible within a few hours or less. The velocity dispersion could be measured with the MIRI spectrograph, as the corresponding fluxes of $2\times10^{-19}-10^{-18}$~erg~s$^{-1}$~cm$^{-2}$ are well above the sensitivity of $2\times10^{-18}$~erg~s$^{-1}$~cm$^{-2}$ for $3\sigma$ detection in $10^4$~s.  

For the H~$\beta$ line at $486.1$~nm, we find that it would be redshifted to $\sim4.4$~$\mathrm{\mu}$m, which is still in the range of NIRCam, with a sensitivity of $6.18$~nJy. Our estimates yield fluxes between $2-20$~nJy, detectable within some hours up to a day. At higher redshift, this will also fall into the MIRI frequency range, where it may be detected even on a shorter timescale due to the smaller intrinsic linewidth of the instrument. 

\begin{table}[htdp]
\begin{center}
\begin{tabular}{lll}
Observable &  $\varphi~[nJy]$ & estimated from \\
\hline
stellar light  & $0.3-3$ & Davies et al. 2007 \\
AGN continuum  & $0.3$ & Spinoglio et al. 2005 \\
Lyman $\alpha$  & $\sim2$ & Dijkstra et al. 2006 \\
H~$\alpha$    &  $100-700$ & Kaspi et al. 2000\\
H~$\beta$   &  $2-20$ & Kaspi et al. 2000 \\
Mg~II~$279.8$~nm  & $10^3-3\times10^3$ & Dietrich et al. 1999  \\
C~IV~$154.9$~nm  & $2\times10^4-10^5$ & Dietrich et al. 2003 \\
\hline
\end{tabular}
\end{center}
\caption{The main observables for JWST, as discussed in \S~\ref{JWST}. We estimate the flux $\varphi$ at $z\sim8$ as described in the text. The NIRCam continuum sensitivities are given in Table \ref{tab:JWST}.}
\label{tab:observablesJWST}
\end{table}%

In addition, one may search for the Mg$^{+}$ line with a rest frame wavelength of $279.8$~nm and the C$^{3+}$ line with $154.9$~nm, which would be redshifted into the NIRCam frequency range and provide complementary information about the velocity dispersion near the black hole. We find that, if the objects from the sample of \citet{Dietrich99} would be shifted to $z\sim8$, the expected fluxes from the C$^{3+}$ line in a typical NIRCam frequency bin (wide filter) would be of the order $1-3$~$\mathrm{\mu}$Jy, and thus detectable within much less than an hour. To estimate the fluxes in the Mg$^{+}$ line, we use the $z=3.4$ sample of \citet{Dietrich03}. We find similarly high fluxes of the order $20-100$~$\mathrm{\mu}$Jy. They may be reduced for less massive black holes, but are well accessible for JWST.

The velocity dispersion of the H~$\beta$, Mg$^{+}$ and C$^{3+}$ lines can be measured with NIRSpec, that provides a sensitivity of $\sim1.5\times10^{-19}$~erg~s$^{-1}$~cm$^{-2}$, while the expected flux is $(1-10)\times10^{-18}$~erg~s$^{-1}$~cm$^{-2}$ for H~$\beta$, $(5-20)\times10^{-16}$~erg~s$^{-1}$~cm$^{-2}$ for C$^{3+}$ and $(1-8)\times10^{-14}$~erg~s$^{-1}$~cm$^{-2}$ for Mg$^+$.

In summary, we find that the component from the starburst will likely be the main observable for JWST, as well as the H~$\alpha$, H~$\beta$, Mg$^{+}$ and C$^{3+}$ lines in case of unobscured broad-line regions. Even the other components alone would be detectable. At redshifts $z\sim5$, typical integration times may reach from a few hours to a day, depending on the luminosity of the source. Near $z\sim10$, typical integration times may vary between one and ten days. The main observables are summarized in Table~\ref{tab:observablesJWST}.

\subsection{The number of high-redshift sources for different observational strategies}

\begin{figure}[t]
\includegraphics[scale=0.4]{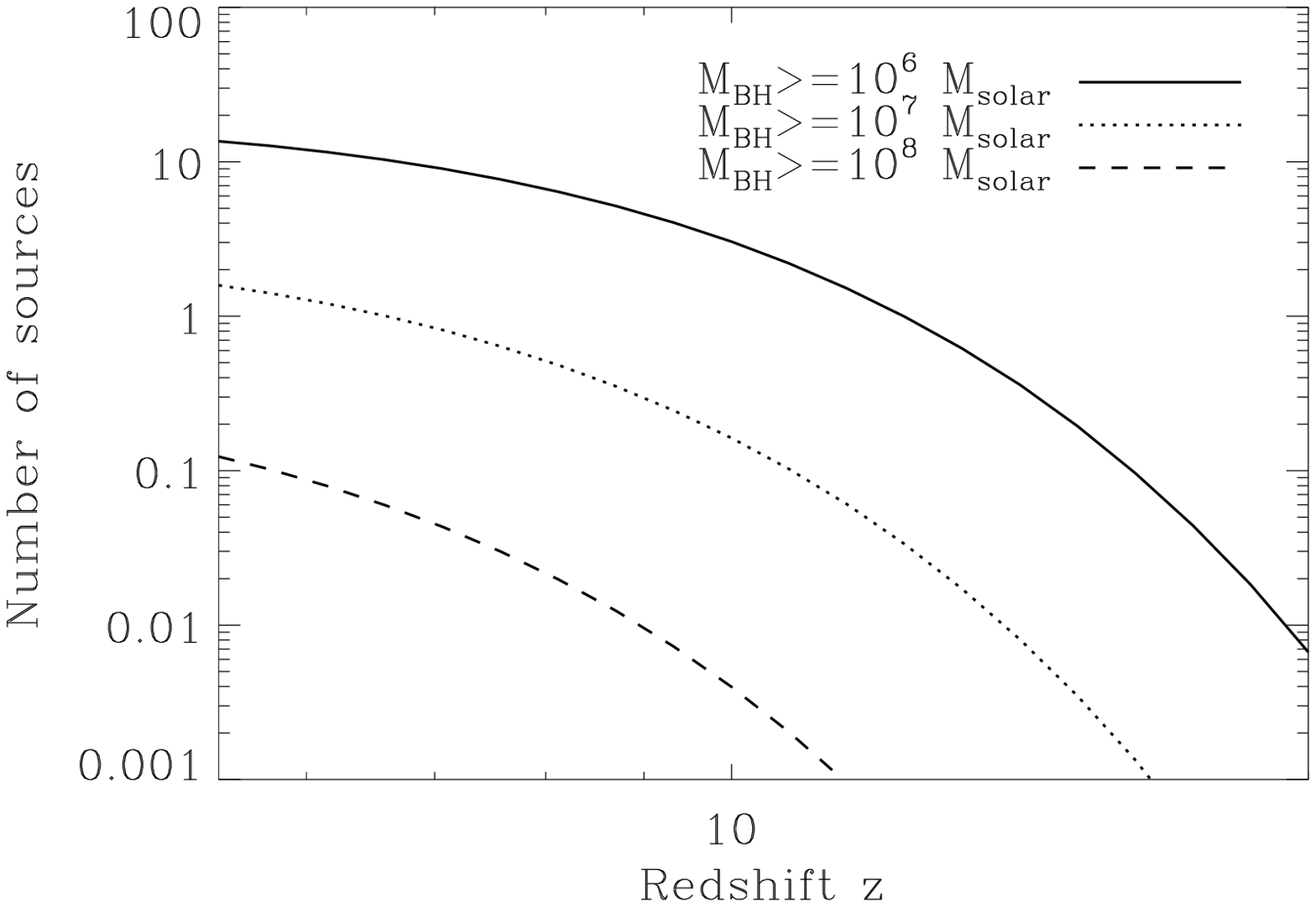}
\caption{The expected number of sources for the first observational strategy focusing on high-$J$~\COI line emission. We adopt a field of view of $(1')^2$, an active quasar fraction of $20\%$ and a redshifter interval of $\Delta z=0.5$ as discussed in the text. Down to $z\sim10$, a typical ALMA field of view therefore has a high chance to find a number of quasars with black holes in the mass range of $10^6-10^7\ M_\odot$. Near $z=6$, one may even find more massive quasars by extending the search to larger volumes. Note that the numbers above scale linearly with the adopted fraction of active quasars.}
\label{fig:strategy1}
\eef

\begin{figure}[t]
\includegraphics[scale=0.4]{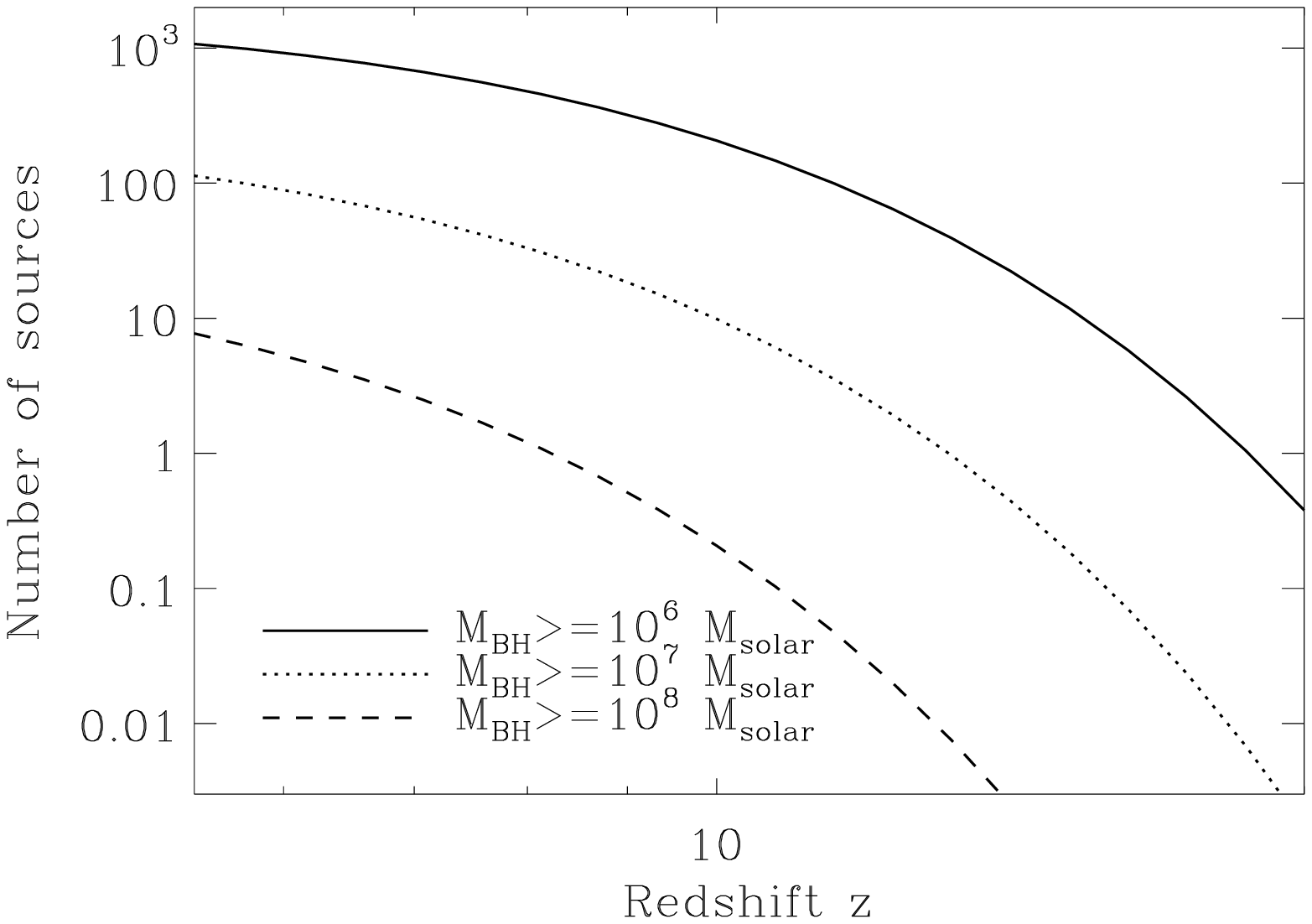}
\caption{The expected number of sources for the second observational strategy focusing on continuum radiation mostly from stellar light. We adopt a field of view of $(2.16')^2$ and an active quasar fraction of $100\%$, as the relevant observables do not require quasar activity. We explore a redshift interval of $\Delta z=2$. In fact, this yields a high number of sources that are potentially accessible to JWST within the NIRCam field of view. For redshifts near $z=5$, typical detection timescales are one day or less, while observations near $z=10$ require integration times between one and ten days. This is within reach of deep surveys.}
\label{fig:strategy2}
\eef

With the considerations above, we can attempt to estimate the number of observable sources at different redshifts. We adopt a similar formalism as \citet{Choudhury07} and count the number of black holes $N(z,M_{\sm{BH}})$ with masses larger than $M_{\sm{BH}}$ in a redshift interval $[z,z+\Delta z]$. This is given as
\begin{eqnarray}
N(z,M_{\sm{BH}})&=&\int_z^{z+\Delta z}dz' \frac{dV}{dz' d\Omega}\int_{M_{\sm{min}}}^{M_{\sm{max}}} d\log M_h \nonumber \\
&\times& f_a \frac{dn}{d\log M_h}(M_{BH}),\label{counts}
\end{eqnarray}
where $dV\ dz'^{-1}\ d\Omega^{-1}$ denotes the comoving volume element per unit redshift per unit solid angle, which is given as \citep{Peebles93}
\begin{equation}
\frac{dV}{dz' d\Omega}=D_A^2 c \frac{dt}{dz}.
\end{equation}
In this expression, $D_A$ is the angular diameter distance, $c$ is the speed of light and $dt/dz=1/\left (H(z)(1+z)\right )$, where $H(z)$ is the expansion rate as a function of redshift. The term $dn/d\log M_h$ describes the number density of dark matter halos per unit mass, and the integration range in Eq.~\ref{counts} is chosen such that the minimal halo mass corresponds to a black hole mass of $M_{\sm{BH}}$ for our model in \S~\ref{population}. The upper halo mass is chosen corresponding to a black hole mass of $10^9\ M_\odot$. The parameter $f_a$ is the fraction of active quasars, as introduced in \S~\ref{constraint}.

As our reference system, we have adopted the known AGN  NGC~1068, which has a black hole of $\sim10^7\ M_\odot$ and a starburst component with $\sim10\ M_\odot$~yr$^{-1}$. As discussed above, even less active quasars provide a sufficient amount of \COI flux for detection, and the same holds for smaller star formation rates. We will therefore assume that typical galaxies containing a $10^6\ M_\odot$ black hole should be observable. To be more conservative, we will also show the corresponding numbers for more massive black holes.

In the following, we will distinguish between two different observational strategies. The first aims at finding high-$J$~\COI lines with ALMA, which will allow us to unambigously identify a given source as a quasar \citep{Spaans08}. The second aims at finding high-redshift sources in general and focuses mainly on continuum emission from dust or stellar light. In fact, once such sources are found, they can be examined in more detail by other observatories like the Keck telescope or new generations of X-ray telescopes.

For the first observational strategy focusing on line observations, ALMA provides a bandwith of $8$~GHz. In general, one may scan a redshift interval of $\Delta z /(1+z)=\Delta \nu/\nu$, where $\Delta\nu$ refers to the difference with respect to the observed frequency $\nu$. For $\nu\sim90$~GHz and $z\sim5$, this corresponds to $\Delta z=0.5$. This becomes even larger when going to higher redshifts, and smaller for observations at higher frequencies, such that we slightly underestimate the sources at high redshift. We adopt this value here as a typical mean. Based on the results in \S~\ref{constraint}, we assume an active quasar fraction of $f_a\sim20\%$. Our results can however be directly scaled to other values. We consider a field of view of $(1')^2$, even though ALMA can probably scan an angular region of $(3')^2$ within a few days. The result is given in Fig.~\ref{fig:strategy1}. 

We find that down to $z\sim10$ there are good chances to detect several sources with black hole masses in the range $10^6-10^7\ M_\odot$ within the ALMA beam. For an extended survey that probes an angular field of $\sim(3')^2$, one may even find quasars with $M_{\sm{BH}}>10^8\ M_\odot$, and at the same time find enough sources with smaller masses to allow a statistical analysis. At redshifts beyond $10$, we find an exponential decrease in the number of high-redshift quasars, though we expect that ALMA could still find some of them. Searching for such systems should be a high priority for future observations, as they are a direct probe for the origin of the first supermassive black holes.

For the second strategy focusing on continuum radiation, we are not restricted to active black holes and work with $f_a=1$. This assumes that such galaxies have a star formation rate of at least $1\ M_\odot$~yr$^{-1}$. However, our results can simply be rescaled for different star formation rates. For continuum observations, the bandwidth is no longer a restriction for the redshift range, and we consider a generic interval of $\Delta z=2$. We adopt the NIRCam field of view of $(2.16')^2$. The result is then given in Fig.~\ref{fig:strategy2}. While we cannot be sure that the detected objects are active quasars, it is always possible to address this question with follow-up observations involving other telescopes. In the NIRCam field of view, we expect $10^2-10^3$ observable sources at redshifts $5-10$, and $10-100$ at redshifts beyond. In addition, we expect a few very bright sources that could be associated with $10^8\ M_\odot$ black holes within the field of view. They can be found with integration times of a few hours, but beyond that a deep field is required to see smaller galaxies and probe the early epoch of galaxy formation in detail.

\subsection{What we will learn about the brightest quasars}
In addition to finding less massive quasars at higher redshift, instruments like ALMA and JWST can be used to get more detailed information about those sources that we already know. At redshift $z=6$, a typical angular resolution of $\sim0.01'$ corresponds to a spatial resolution of $\sim60$~pc. It is thus possible to resolve the quasar host galaxy in high-$J$~\COI emission and dust with ALMA. As these sources are very bright in X-ray emission, the required flux to density ratio that produces significant \COI emission is shifted to larger scales further away from the central black hole and fills a larger region in space. With detailed diagnostics for the \COI lines, it is thus possible to probe temperature and density in these systems with spatial resolution. With a spectral resolution of up to $0.01$~km~s$^{-1}$, the dynamics of the gas can be probed in detail, in particular with respect to typical infall velocities as well as winds and jets.

In addition, JWST will provide complementary information about stellar light, the AGN continuum and Lyman $\alpha$ emission. The latter could be clearly discriminated from the continuum contribution using specific narrow filters tuned to the wavelength corresponding to the redshifted Lyman $\alpha$ line.

\section{Discussion and outlook}\label{discussion}
In this paper, we have presented a model for the quasar population at high redshift, based on the known correlations between the black hole mass and its host galaxy. We have derived constraints on the fraction of active quasars from the SXRB and discussed the expected observables. We have based our discussion on using NGC~1068 as a template AGN, and calculated its observability at high redshift. Our focus lies in particular on high-$J$~\COI lines, fine-structure lines of \CII and \OId, dust emission, stellar light, the AGN continuum and Lyman $\alpha$ radiation. Above a metallicity of $10^{-2}$ solar, \COId, \CII and \OI are the main coolants, and their detection can be used to derive an estimate for the heating rate and the impinging X-ray flux. A profile of the observed line width further allows to derive estimates for the black hole mass and probes the mass - luminosity relation at high redshift. We expect that galaxies containing a black hole in the mass range $10^6-10^7\ M_\odot$ can be observed at high redshift with ALMA and JWST. We predict $1-10$ high-redshift sources that can be found via molecular lines in an ALMA field of view of $(1')^2$, and $100-1000$ sources to be found by JWST using NIRCam and focusing on stellar emission.

One uncertainty in this discussion comes from the question whether the Maggorian relation is established already at high redshift. This seems fair in light of supersolar metallicities in high-redshift quasars. On the other hand, if our model overpredicts the number of high-redshift black holes, the X-ray constraints would be alleviated further, and a larger fraction of the black hole population could be active. In this sense, the number of active quasars might still be of the same order. Indeed, unless the population of black holes between redshifts $5$ and $6$ has accreted mass with high super-Eddington rates, there must be progenitors with masses of $\sim10^6-10^7\ M_\odot$ at higher redshift. It is a secondary issue for our purposes whether the black hole would be directly visible in X-rays, or if the system is a hidden AGN and only visible in the excited high-$J$~\COI lines.

In about three years, ALMA will become operational with initially $16$~antennas. This will provide sufficient sensitivity to detect high-$J$~\COI lines, fine-structure lines and dust continuum emission from the brightest high-redshift quasars that are already known. As these systems are large, the high angular resolution of ALMA will provide detailed spatial information about the inner structure of such systems, and detailed diagnostics for the \COI lines allow to infer the density and temperature distribution in the host galaxy \citep{Meijerink07}. 

Once ALMA has the full set of $50$~antennas, it reaches even higher sensitivity and it becomes possible to search for AGNs with less massive black holes. For such deep field observations, we recommend to focus initially on modest redshifts of $z\sim5-6$, where we expect higher fluxes, more sources, and where one may be more confident that the Magorrian relation is well-established. This provides an important test to confirm the observability of high-redshift quasars, as well as important information about the progenitors of the most massive quasars in the universe. Such studies can then gradually be extended to higher redshifts and will provide a direct test for structure formation models in the high-redshift universe and of the Magorrian relation at different redshifts.

JWST can aid such surveys by providing additional sources that may be detected in stellar light, the AGN continuum or Lyman $\alpha$ emission. In this case, ALMA does not need to glance into the dark sky, but can do pointed follow-ups of the known sources. This will speed up high-redshift surveys, and provide complementary information at different frequencies. It is particularly important at very high redshift beyond $z\sim10$, where the probability of finding sources in the ALMA field of view decreases rapidly. The combination of these instruments can therefore truely detect the first quasars in the universe.

\acknowledgments

We thank Robi Banerjee, Andrea Ferrara, Wilfred Frieswijk, Simon Glover, Thomas Greif, Edo Loenen, Rowin Meijerink and Dieter Poelman for valuable discussions on the topic. DRGS thanks the Heidelberg Graduate School of Fundamental Physics (HGSFP), the LGFG and the Graduate Academy Heidelberg for financial support. The HGSFP is funded by the Excellence Initiative of the German Government (grant number GSC 129/1). RSK thanks for support from the Emmy Noether grant KL 1358/1. DRGS and RSK also acknowledge subsidies  from the DFG SFB 439 {\em Galaxies in the Early Universe}. 


\clearpage




\clearpage

\end{document}